\documentstyle[prb,aps,graphicx]{revtex}

\begin{document}
\draft

\wideabs{
\title{Vortex Matter Transition in
Bi${}_2$Sr${}_2$CaCu${}_2$O${}_{8+y}$ under Tilted Fields }

\author{S. Ooi,$^{1,\ast}$ T. Shibauchi,$^{1,\dag}$ 
K. Itaka,$^{1}$ N. Okuda,$^{1}$ and T. Tamegai$^{1,2}$}
\address{$^1$Department of Applied Physics, The University of Tokyo,
7-3-1 Hongo, Bunkyo-ku, Tokyo 113-8656 Japan}
\address{$^2$CREST, Japan Science and Technology Corporation (JST)}

\date{\today}
\maketitle
\begin{abstract} 
Vortex phase diagram under tilted fields from the $c$ axis
in Bi${}_2$Sr${}_2$CaCu${}_2$O${}_{8+y}$
is studied by local magnetization hysteresis measurements using Hall probes.
When the field is applied at large angles from the $c$ axis,
an anomaly ($H_p^\ast$) other than the well-known 
peak effect ($H_p$) is found at fields below $H_p$.
The angular dependence of the field $H_p^\ast$
is nonmonotonic and clearly different from that of $H_p$
and depends on the oxygen content of the crystal.
The results suggest existence of a vortex matter transition
under tilted fields.
Possible mechanisms of the transition are discussed.
\end{abstract}

\pacs{PACS numbers: 74.60.Ec, 74.72.Hs, 74.25.Dw, 74.25.Ha}
}

It has been made clear that
the vortex state in high-$T_c$ superconductors has
various interesting features.\cite{Blatter94A}
Important parameters which control these features
are large thermal fluctuations due to high critical temperature $T_c$
and the anisotropy of these materials.
Although the anisotropy parameter $\gamma (\equiv \lambda_c/\lambda_{ab})$
is very different in two typical high-$T_c$ superconductors,
YBa$_2$Cu$_3$O$_{7-\delta}$\ (YBCO) $\gamma \sim 5$ (ref.2)
and
Bi${}_2$Sr${}_2$CaCu${}_2$O${}_{8+y}$\ (BSCCO) $\gamma \sim 200$
(ref.3),
similar phenomena such as
the first-order phase transition (FOT) \cite{Safer92,Schilling96,Zeldov95}
and the (second) peak effect \cite{Tamegai93,Deligiannis97}
are commonly observed.
The difference of anisotropy is reflected in the fact
that these transitions in BSCCO occur in a much lower-field range than
those in YBCO.
An interesting question one may ask is whether the difference of anisotropy
induces a qualitative difference in the vortex states or not.

When the field is inclined from the $c$ axis,
a drastic change in the vortex order is expected due to large anisotropy.
A vortex chain state in which all vortices form arrays in the tilt plane
has been observed
under tilted fields in YBCO by the Bitter decoration method.\cite{Gammel92}
The existence of such a vortex chain in anisotropic superconductors
has been predicted
within the framework of the anisotropic London model.\cite{Buzdin90}
The motive force of the chain formation originates
from the attractive magnetic interaction between inclined vortices.
On the other hand,
vortices form more complicated arrangements in BSCCO.
Vortex chains embedded in approximately triangular vortex lattices
have been observed.\cite{Bolle91,Grigorieva95}
This vortex ordering has attracted theoretical attention and
the coexistence of two vortex species
has been suggested.\cite{Huse92,Daemen93b,Preosti93}
Experimentally, Grigorieva {\it et al.} concluded
that vortex chains embedded in the vortex lattice consist
of two inclined vortex species
which have different angles from the $c$ axis.\cite{Grigorieva95}
These results suggest that
the difference of anisotropy between YBCO and BSCCO
causes different vortex arrangements under tilted fields.

Such observations of vortex states under tilted fields, however,
have been limited to very low fields ($<$ 200 Oe).
Our purpose is to clarify the vortex phase diagram in BSCCO single crystals
under tilted applied fields in a wide field range ($<$10\ kOe).
In this paper,
we report on an anomaly in the local magnetization in BSCCO
which is observed only under tilted fields.
This anomaly suggests the existence of vortex matter transition
under tilted fields.
Based on the angular, temperature, 
and doping-level dependence of the anomaly,
we discuss the origin of the transition.

\begin{figure}
\includegraphics[width=0.8\linewidth]{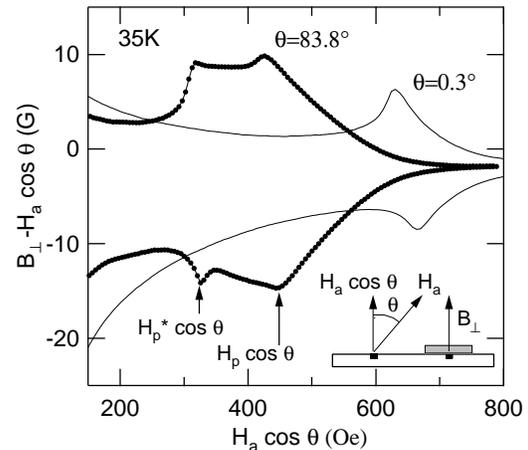}
\caption{
Local magnetization hysteresis curves at $\theta=0.3^\circ$ 
and $\theta=83.8^\circ$ in the OV1 sample.
Arrows indicate the positions of two characteristic anomalies.
The horizontal axis shows
the field parallel to the $c$ axis, $H_a \cos \theta$.
The inset shows the configuration of a sample 
and micro-Hall probes schematically.
}
\end{figure}

Single crystals of BSCCO have been grown by the floating-zone method .
\cite{Ooi98a}
We measured several samples with different oxygen contents,
which were controlled by annealing at 300-500$^\circ$C
for one day in appropriate partial oxygen pressure.
In this paper, we show results for four samples;
optimally doped (OPT), overdoped (OV1, OV2), and
highly overdoped (HOV) crystals.
The critical temperatures $T_c$ are
86.7, 80.3, 83.5, and 76.8 K, respectively.
The configuration of the measurement by using micro-Hall probes 
is schematically shown in the inset of Fig.1.
The details of the measurement are described 
in the previous paper.\cite{Ooi99b}


A typical local magnetization curve at 35 K 
under a tilted field from the $c$ axis
is shown with a curve in the field parallel to the $c$ axis in Fig.\ 1.
In addition to the well-known peak effect ($H_p$) \cite{Tamegai93},
an anomaly, indicated by $H_p^\ast$, is observed below $H_p$.
This anomaly is observed only under tilted fields.
We confirmed the presence of this anomaly
in more than ten single crystals and Pb-doped BSCCO crystals.\cite{Itaka99}

Figures 2(a) and 2(b) show the first quadrant part of
the local magnetization hysteresis curves at 35\ K
in the OV1 sample at various angles from 0$^\circ$ to 84$^\circ$.
When the field is nearly parallel to the $c$ axis,
only the peak effect ($H_p$) is observed at 630 Oe.
On the other hand,
the magnetization anomaly $H_p^\ast$ starts to be observed below $H_p$
at angles above about 60$^\circ$ in this crystal.
Since the anomalies have a broad steplike forms at lower angles,
the field $H_p^\ast$ 
is picked up as the position of the shoulder of magnetization,
which is shown by the dashed line in Fig.\ 2.
At larger angles, the anomaly has sharper shape
and the definition of $H_p^\ast$ is not ambiguous.

\begin{figure}[t]
\begin{center}
\includegraphics[width=0.8\linewidth]{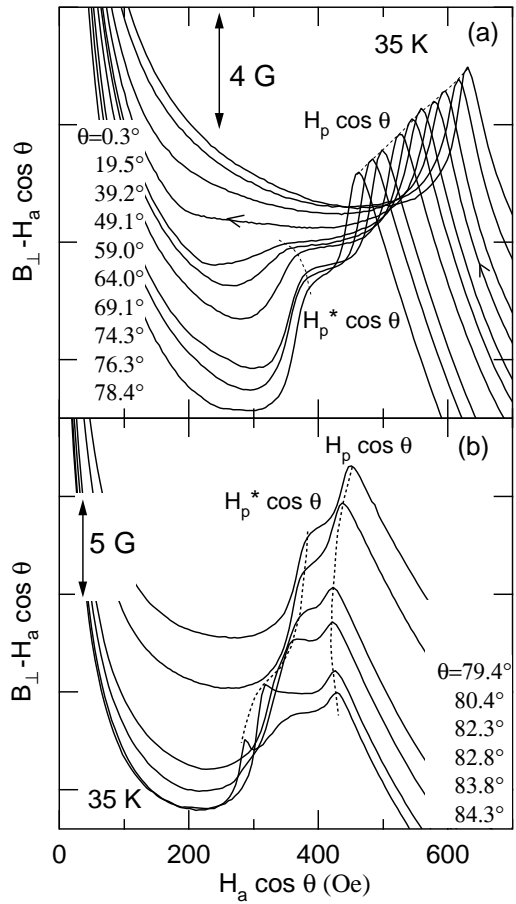}
\end{center}
\caption{
First quadrant part of hysteresis curves at 35\ K in the OV1 sample
at angles (a) below $\theta^\ast$
and (b) around $\theta^\ast$.
Dashed lines indicate peak positions $H_p \cos \theta$ and
$H_p^\ast \cos \theta$.
The definition of $\theta^\ast$ is described in Fig.\ 3.
Each curve is properly shifted along the vertical axis for clarity.
}
\end{figure}

\begin{figure}[t]
\begin{center}
\includegraphics[width=0.8\linewidth]{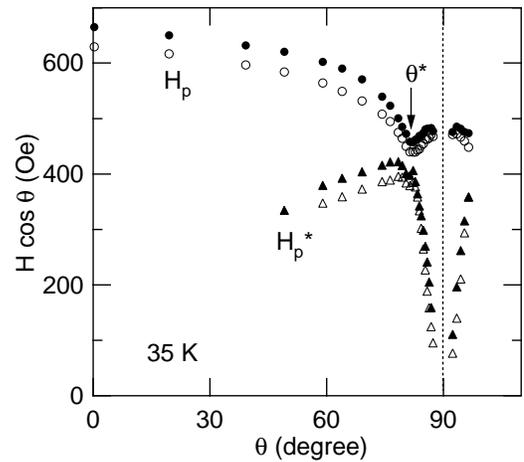}
\end{center}
\caption{
Angular dependence of $H_{p}^\ast \cos \theta$ (triangles) and
$H_{p} \cos \theta$ (circles) in the OV1 sample at 35\ K.
Closed and open markers show the peak fields 
in field-increasing and -decreasing processes, respectively.
At $\theta^\ast$, the angular dependence of $H_{p} \cos \theta$
has a minimum.
}
\end{figure}

The angular dependence of $H_p^\ast \cos \theta$ is plotted in Fig.\ 3
together with the normal peak effect.
The difference of peak positions
between increasing and decreasing field process
is due to the self-trapped field in the crystal.
The angular dependence of $H_p$ follows the scaling relation,
$H_p \propto (\cos \theta + \alpha \sin \theta)^{-1}$,
as reported previously.\cite{Ooi99b}
At angles higher than a characteristic angle $\theta^\ast$,
$H_p \cos \theta$ deviates from the scaling and starts to increase.
On the other hand, 
$H_p^\ast \cos \theta$ slightly increases with increasing angle
up to $\theta^\ast$, and it decreases rapidly above $\theta^\ast$.
At around $\theta^\ast$
both $H_p \cos \theta$ and $H_p^\ast \cos \theta$ 
approach closely to each other.
Since the angular dependence of $H_p^\ast$
is quite different from that of $H_p$,
the origins for these two anomalies should be different. 

Figure\ 4(a) shows the hysteresis curves at 45 and 55\ K
at a constant angle of 83.7$^\circ$ which is larger than $\theta^\ast$.
The normal peak effect observed at 45\ K and 630\ Oe
changes into the magnetization step at 55\ K which is a sign of the FOT.
In addition, a steplike structure is observed
at a field lower than $H_p$.
This anomaly can be traced to the anomaly $H_p^\ast$
observed at lower temperatures as shown in Fig.\ 1.
We could not observe an anomaly
within our experimental resolution $\sim 10 $mG 
when the irreversible magnetization disappears at high temperatures.
The temperature dependence of $H_{FOT}$, $H_p$, and $H_p^\ast$
at 83.7$^\circ$ is shown in Fig. 4(b).
One can see that the $H_p^\ast(T)$ line is always 
below the normal transition lines ($H_{FOT}$, $H_p$) even above $\theta^\ast$,
which rules out a possibility that $H_p(\theta)$ crosses $H_p^\ast(\theta)$
at $\theta^\ast$ (see Fig.\ 3). 
Thus this $H_p^\ast$ line divides the vortex lattice phase
into two parts under an oblique field.

\begin{figure}[t]
\begin{center}
\includegraphics[width=0.75\linewidth]{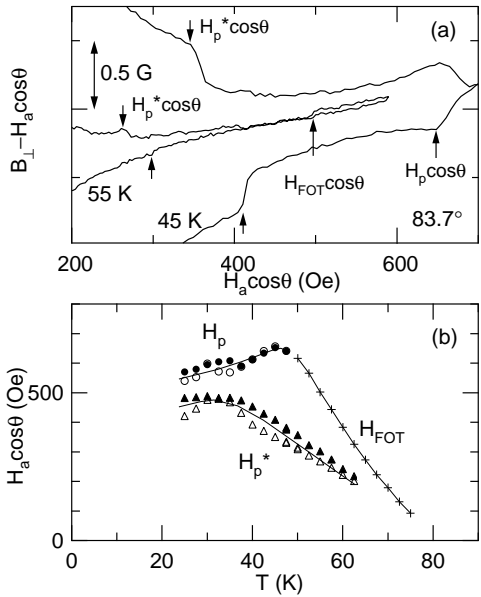}
\end{center}\caption{
(a) Hysteresis curves at 45 and 55\ K at a constant angle of 83.7$^\circ$
in the OV2 sample.
Arrows indicate positions of the anomalies.
(b) Temperature dependence of 
the $c$-axis component of $H_{FOT}$, $H_p$, and $H_p^\ast$.
Closed and open markers indicate the field increasing and decreasing processes,
respectively.
}
\end{figure}

Let us discuss possible origins of the observed anomaly at $H_p^\ast$.
Dynamical scenario may also be applied in this case as in the case of 
the fish-tail effect in YBCO.\cite{Krusin-Elbaum}
However, in such a case, changes in the irreversible magnetization is 
smoother compare with the present case at $H_p^\ast$.
The situation is rather similar to the case of the peak effect in BSCCO 
,\cite{Tamegai93}
where the irreversible magnetization shows a steplike change.
The peak effect is believed to be a transition in the vortex solid phase,
and is accompanied by a drastic change 
in the vortex configuration.\cite{Ertas96,Horovitz98}
A disorder-induced transition \cite{Ertas96,Horovitz98} and 
a dimensional crossover of the vortex system \cite{Tamegai93}
are proposed as possible origins of the peak effect.
Since the critical current is related to the collective pinning force 
determined by the configuration of vortices and the pinning potential, 
a change in the vortex configuration induce a steplike change of 
the critical current at the characteristic field. 
As a result, the irreversible magnetization show a distinct anomaly, 
as in the present case at $H_p^\ast$.
Although magnetization step is not observed in the reversible regime,
the height of the step might be simply too small to be observed.
Since the transition occurs within the vortex solid state, 
it is reasonable that the entropy change, 
which is proportional to the magnetization step, is much smaller
than the case of the vortex lattice melting transition.

Several theories predict an existence of a transition of the vortex arrangements in highly anisotropic layered superconductors under tilted fields.
A recent theoretical work by Koshelev \cite{Koshelev99} showed that 
the ground state under the tilted fields except for $\theta \sim 0^\circ$
is the crossing lattice state consisting of
almost independent units of pancake and Josephson vortices 
in such superconductors.
The interesting Bitter decoration pattern of vortex chains
embedded in vortex lattices observed in BSCCO single crystals \cite{Bolle91}
is explained by the mechanism that Josephson vortices attract
additional pancake vortices.
In the same framework, 
experimentally observed linear dependence of $H_{FOT} \cos
\theta$ on the in-plane component of field \cite{Ooi99b}
is also understood. 
In this theory,
a transition from the mixed chains-lattice state to the distorted lattice
is suggested in the crossing lattice state.
This transition is a good candidate for the origin 
of the anomaly at $H_p^\ast$.
Experimental data below $\theta^\ast$ can be fitted 
by the estimated transition field (Eq.\ 11 of Ref.\ 22) as shown in Fig.\ 5(a).
Although we have a reasonable fitting to the data in $\theta<\theta^\ast$ and
the reasonable values of $\gamma$
(600, 240, 200, 140 for the OPT, OV1, OV2, HOV samples, respectively), 
the obtained penetration depth $\lambda=400$\ \AA\  is too short compared with
generally accepted value of about 2000\ \AA\ in BSCCO.
Strictly speaking, the formula is valid 
when pancake-lattice spacing $a \gg \lambda$ and $\gamma \gg \lambda/s$,
which do not hold at $H_p^\ast$.
However, general trend is believed to be valid beyond this range.
This might be the reason why we get unexpectedly short $\lambda$ from the
fitting.

\begin{figure}[htbp]
\begin{center}
\includegraphics[width=0.75\linewidth]{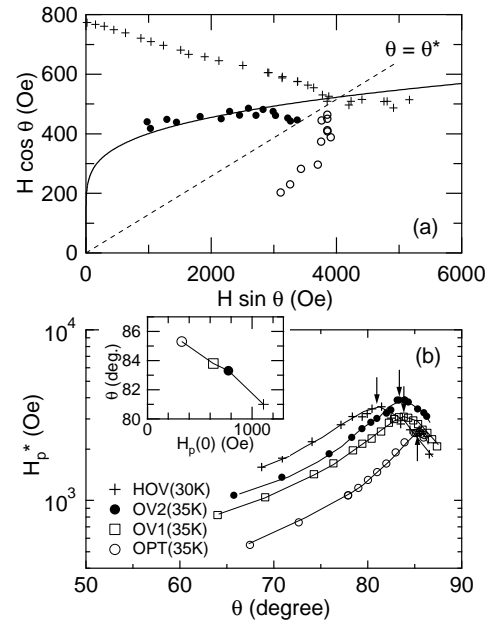}
\end{center}\caption{ 
(a) Dependence of $H_p^\ast \cos \theta$
on the in-plane component of the field 
for the OV2 sample below (closed circles) 
and above (open circles) $\theta^\ast$.
Additionally, $H_p \cos \theta$ (pluses) is also plotted.
Solid curve shows a fit to the transition line from
the mixed chains-lattice state to  the distorted lattice
with $\gamma=200$, $\lambda=400$\AA\  and 
the distance between neighboring superconducting layers $s$=15\AA.
(b) Angular dependence of $H_p^\ast$ in decreasing-field branch 
for four samples with different doping levels.
Solid lines are guides to the eye.
Arrows indicate the angle where $H_p^\ast(\theta)$ shows a maximum.
This angle is plotted as a function of $H_p(0)$ in the inset.
}
\end{figure}


In this scenario,
deviation from the above behavior at angles larger than $\theta^\ast$ 
might reflect 
the presence of additional transition in the vortex lattice structure.
Particularly, it is interesting to point out 
that the transition in this angle range occurs
at an almost constant in-plane field value as shown in Fig. 5(a), 
suggesting the contribution from the Josephson vortex lattice.
The deviation from the scaling law of $H_p$ above $\theta^\ast$
may also be understood as reflecting a change in Josephson vortex lattice,
such as the overlap of Josephson vortices.

Another possible transition is the one from the mixed chains-lattice state
to uniformly tilted lattice state.
Bitter decoration pattern observed by Bolle {\it et al}. \cite{Bolle91} 
implies that the uniformly tilted lattice
is not an equilibrium phase in low and oblique fields
in superconductors with large anisotropy.
Besides Koshelev's theory, 
the instability of uniformly tilted vortex lattice has been predicted
by several theories.\cite{Sardella93,Nguyen96,Thompson97}
Particularly, Thompson and Moore \cite{Thompson97} have suggested that
uniformly tilted vortex lattice state
including vortex chain state observed in YBCO
is not stable in low and tilted fields,
when the superconductor has larger anisotropy ratio than a threshold.
According to their calculation,
the threshold is $\gamma \sim 12$ for $\kappa = 50$,
which is much smaller than $\gamma$ of BSCCO.

To compare the anomaly with the instability of the vortex lattice,
the angular dependence of $H_p^\ast$ in four samples with different
oxygen contents is plotted in Fig.\ 5(b).
The characteristic features that $H_p^\ast(\theta)$
reaches maximum at about 80$^\circ$ and
decreases at higher angles
are consistent with the expectation of the instability field
for $\gamma=60$ and $\kappa=50$
by Thompson and Moore.\cite{Thompson97}
The angle where the instability field has a maximum is predicted to shift 
to lower angle at smaller anisotropy $\gamma$.
This is actually observed in our experiments.
As shown in the inset of Fig.\ 5(b),
this angle decreases with decreasing anisotropy,
which is known to be inversely correlated to the peak field $H_p(0)$
when $H \parallel c$.\cite{Kishio94a}
Another point is that the temperature dependence of the instability field
is expected to be proportional to 
the mean-field upper critical field $B_{c2}(T)$,
which may explain the observed linear temperature dependence 
above 0.5$T_c$ [Fig.\ 4(b)].
As shown above, qualitative features of the 
magnetization anomaly $H_p^\ast$
are consistent with the calculation of Thompson and Moore.

In summary,
we measured the local magnetization hysteresis curves
under tilted fields from the $c$ axis in BSCCO single crystals.
We found a magnetization anomaly in the vortex lattice phase,
which is not present when the field is parallel to the $c$ axis.
Observed sharp change of magnetization at the anomaly ($H_p^\ast$)
suggests the change of the vortex ordering as a possible origin of
the anomaly. 
Transition from the mixed chains-lattice state to the distorted
lattice state as suggested recently by Koshelev may explain this anomaly.
Alternatively, the angular, temperature, and oxygen doping dependence
of $H_p^\ast$
are qualitatively similar to the theoretical calculation for
the instability of the uniformly tilted vortex lattice. 
Direct observations of vortex lattice ordering 
under tilted fields in the field range 
of interest are highly desirable.

The authors are grateful to A. E. Koshelev for stimulating discussion.
This work is supported by Grant-in-Aid for Scientific Research
from the Ministry of Education, Science, Sports and Culture.
S. O. acknowledges the supports by JSPS
for Research Fellowships for Young Scientists.




\end{document}